\documentclass[3p,times,twocolumn]{elsarticle}

\usepackage{ecrc}

\volume{00}

\firstpage{1}

\journalname{Nuclear and Particle Physics Proceedings}

\runauth{B. Duclou\'e, T. Lappi, H. M\"antysaari}

\jid{nppp}

\jnltitlelogo{Nuclear and Particle Physics Proceedings}

\usepackage{amssymb}
\usepackage[figuresright]{rotating}

\usepackage[utf8]{inputenc}
\usepackage{amsmath}
\usepackage{hyperref}

\def\P{{\boldsymbol P}}
\def\k{{\boldsymbol k}}
\def\l{{\boldsymbol l}}
\def\p{{\boldsymbol p}}
\def\q{{\boldsymbol q}}

\newcommand{\der}{\mathrm{d}}
\newcommand{\xt}{{{\boldsymbol x}_\perp}}
\newcommand{\yt}{{{\boldsymbol y}_\perp}}
\newcommand{\bt}{{{\boldsymbol b}_\perp}}
\newcommand{\rt}{{{\boldsymbol r}_\perp}}
\newcommand{\kt}{{\k_\perp}}
\newcommand{\lt}{{\l_\perp}}
\newcommand{\pt}{{\p_\perp}}
\newcommand{\qt}{{\q_\perp}}

\newcommand{\ud}{\, \mathrm{d}}
\newcommand{\tr}{\, \mathrm{Tr} \, }
\newcommand{\nc}{{N_\mathrm{c}}}
\newcommand{\da}{d_\mathrm{A}}

\newcommand{\qso}{Q_\mathrm{s0}}

\newcommand{\lqcd}{\Lambda_{\mathrm{QCD}}}
\newcommand{\as}{\alpha_{\mathrm{s}}}

\newcommand{\Jpsi}{{J/\psi}}

\begin{document}

\begin{frontmatter}



\dochead{}

\title{Forward $J/\psi$ and $D$ meson nuclear suppression at the LHC}


\author[jyu,hip]{B. Duclou\'e}
\author[jyu,hip]{T. Lappi}
\author[bnl]{H. M\"antysaari}
\address[jyu]{Department of Physics, University of Jyv\"askyl\"a, P.O. Box 35, 40014 University of Jyv\"askyl\"a, Finland}
\address[hip]{Helsinki Institute of Physics, P.O. Box 64, 00014 University of Helsinki, Finland}
\address[bnl]{Physics Department, Brookhaven National Laboratory, Upton, NY 11973, USA}

\begin{abstract}
Using the color glass condensate formalism, we study the nuclear modification of forward $J/\psi$ and $D$ meson production in high energy proton-nucleus collisions at the LHC. We show that relying on the optical Glauber model to obtain the dipole cross section of the nucleus from the one of the proton fitted to HERA DIS data leads to a smaller nuclear suppression than in the first study of these processes in this formalism and a better agreement with experimental data.
\end{abstract}


\end{frontmatter}

% \linenumbers

\section{Introduction}

The study of open and hidden charm production in high energy proton-proton and proton-nucleus collisions can be an important probe of gluon saturation. Indeed, the measurement of these processes allows to reach very small $x$ values where saturation effects should be enhanced. Such measurements have been performed at the LHC, in particular by the ALICE and LHCb collaborations. On the theoretical point of view, the charm quark mass should be large enough to provide a hard scale, allowing the use of perturbative techniques, but still small enough to be sensitive to gluon saturation.

Here we will study the nuclear modification of $\Jpsi$ and $D$ meson production in minimum bias collisions at the LHC (we refer to Refs.~\cite{Ducloue:2015gfa,Ducloue:2016pqr} for more detailed studies).
For this we will use the dilute-dense limit of the color glass condensate (CGC) framework, since forward particle production probes the projectile at rather large $x$ and the target at very small $x$. In this approach, the physical representation of the process is that of a collinear gluon emitted by the the projectile proton which can split into a $c\bar{c}$ pair either before or after scattering off the target. The gluon or $c\bar{c}$ pair propagating in the target is then assumed to interact eikonally with it, picking up either an adjoint or fundamental Wilson line factor depending on the particle. This process can then be described in terms of the same Wilson line correlators which appear in other processes, such as (inclusive and diffractive) DIS, single and double inclusive hadron production in proton-proton and proton-nucleus collisions and the initial state for the hydrodynamical modeling of heavy ion collisions. This framework can thus be applied to a broad range of processes.

The nuclear modification of $\Jpsi$ and $D$ meson production has been studied in this formalism in the past~\cite{Fujii:2013gxa,Fujii:2013yja}. However it was observed later at the LHC that the nuclear suppression of forward $\Jpsi$ production was significantly smaller than the one predicted by this calculation. Here we study these processes using the same ``hybrid'' framework, but with a more careful treatment of nuclear geometry when going from a proton to a nucleus target.
Our main motivation is that it was shown~\cite{Lappi:2013zma} that nuclear geometry effects can explain the disagreement between early CGC calculations~\cite{Albacete:2010bs} and LHC data in single inclusive hadron production.

\section{Formalism}

In this work we will focus on the nuclear modification factor, in which normalization uncertainties cancel. This ratio is defined as
\begin{align}\label{eq:defrpa}
R_\text{pA}= \frac{1}{A}\frac{\left . \ud \sigma/ \ud^2 
	\P_\perp \ud Y \right |_\text{pA}}
{\left . \ud\sigma/\ud^2 \P_\perp \ud Y \right |_\text{pp}} \; ,
\end{align}
where $\left . \ud\sigma/\ud^2 \P_\perp \ud Y \right |_\text{pp}$ and $\left . \ud \sigma/ \ud^2 
\P_\perp \ud Y \right |_\text{pA}$ are the cross sections in proton-proton and proton-nucleus collisions respectively. The key quantity needed to obtain the cross sections for $\Jpsi$ and $D$ meson production in proton-proton and proton-nucleus collisions is the cross section for $c\bar{c}$ pair production. Gluon and quark pair production has been studied in great detail in Refs.~\cite{Blaizot:2004wu,Blaizot:2004wv} (see also Ref.~\cite{Kharzeev:2012py}) and applied to various processes, such as~\cite{Fujii:2005rm,Fujii:2006ab,Fujii:2013gxa,Fujii:2013yja}. We use the collinear approximation for the projectile proton since we consider forward rapidities, where it is probed at rather large $x$. We will use the MSTW 2008~\cite{Martin:2009iq} LO parametrization for this. In this approximation, the $c\bar{c}$ pair production cross section reads, in the large-$\nc$ limit~\cite{Fujii:2013gxa}:
\begin{multline}
\!\!\!\!\! \frac{\ud \sigma_{c\bar{c}}}{\ud^2\pt \ud^2\qt \ud y_p \ud y_q}
= 
\frac{\as^2 \nc}{8\pi^2 \da}
\frac{1}{(2\pi)^2}
\!\!\int\limits_{\k_\perp}\!
\frac{\Xi_{\rm coll}(\pt + \qt,\k_{\perp})}{(\pt + \qt)^2}
\\ \times
\phi_{y_2=\ln{\frac{1}{x_2}}}^{q\bar{q},g}(\pt + \qt,\k_\perp)
\;
x_1 G_p(x_1,Q^2) \; ,
\label{eq:dsigmaccbarcoll}
\end{multline}
where $\pt$ and $\qt$ are the transverse momenta of the quarks, $y_p$ and $y_q$ their rapidities, $\int_{\k_\perp} \equiv \int \ud^2 \k_\perp / (2\pi)^2$ and $\da\equiv \nc^2-1$. The expression for $\Xi_{\rm coll}$ is given in Ref.~\cite{Fujii:2013gxa}. The function $\phi^{q \bar{q},g}$, which describes the propagation of the $c\bar{c}$ pair in the color field of the target, reads
\begin{equation}\label{eq:defphi}
\phi_{_Y}^{q \bar{q},g}(\lt,\kt)=
\int\der^2 \bt \frac{N_c \, \l^2_\perp}{4 \alpha_s} \; 
S_{_Y}(\kt) \;
S_{_Y}(\lt-\kt) \;,  
\end{equation}
where $\bt$ is the impact parameter. Here $S_{_Y}(\kt)$ is the Fourier transform of the fundamental representation dipole correlator $S_{_Y}(\rt)$ of the target, with
\begin{equation}
S_{_Y}(\xt-\yt) = \frac{1}{\nc }\left< \tr U^\dag(\xt)U(\yt)\right>,
\end{equation}
and $U(\xt)$ is a Wilson line in the fundamental representation.

In this work $S_{_Y}(\rt)$ is obtained by solving numerically the running coupling Balitsky-Kovchegov equation~\cite{Balitsky:1995ub,Kovchegov:1999ua,Balitsky:2006wa}. The initial condition for the evolution involves non-perturbative dynamics and can be fitted to data. In the case of a proton target we use the MV$^e$ parametrization introduced in Ref.~\cite{Lappi:2013zma}: at the initial rapidity, we have
\begin{equation}\label{eq:icp}
S_{Y= \ln \frac{1}{x_0}}(\rt) = \exp \left[ -\frac{\rt^2 \qso^2}{4} \ln \left(\frac{1}{|\rt| \lqcd}\!+\!e_c \cdot e\right)\right] ,
\end{equation}
with $x_0=0.01$. The running coupling in coordinate space is
\begin{equation}
\as(r) = \frac{12\pi}{(33 - 2N_f) \log \left(\frac{4C^2}{r^2\lqcd^2} \right)} \; .
\end{equation}
A fit to HERA DIS data~\cite{Aaron:2009aa} at $Q^2<50$ GeV$^2$ and $x<0.01$ leads to $\qso^2= 0.060$ GeV$^2$, $C^2= 7.2$, $e_c=18.9$  and $\sigma_0/2 = 16.36$ mb~\cite{Lappi:2013zma}. In this model there is no impact-parameter dependence in the case of a proton target and $\sigma_0/2$ corresponds to the effective transverse area of the proton. Therefore when computing proton-proton cross sections we make the replacement $\int\der^2 \bt \to \sigma_0/2$ in~(\ref{eq:defphi}).

Because of the lack of nuclear DIS data in a kinematical domain similar to the one explored in~\cite{Aaron:2009aa}, this procedure cannot be applied when the target is a nucleus. In Refs.~\cite{Fujii:2013gxa,Fujii:2013yja} the initial condition for the BK evolution of the nucleus was taken of the same form as for a proton but with an initial saturation scale scaled by a factor $\sim A^{1/3}$. Instead, in this work we use the optical Glauber model. In this model, the initial condition for a nucleus reads~\cite{Lappi:2013zma}
\begin{multline}\label{eq:ica}
S^A_{Y=\ln \frac{1}{x_0}}(\rt,\bt) = \exp\bigg[ -A T_A(\bt) 
\frac{\sigma_0}{2} \frac{\rt^2 \qso^2}{4} 
\\ \times
\ln \left(\frac{1}{|\rt|\lqcd}+e_c \cdot e\right) \bigg] \; ,
\end{multline}
where $T_A$ is the standard transverse thickness function given by
\begin{equation}
T_A(\bt)= \int dz \frac{n}{1+\exp \left[ \frac{\sqrt{\bt^2 + z^2}-R_A}{d} \right]} \; ,
\end{equation}
with $d=0.54\,\mathrm{fm}$ and $R_A=(1.12A^{1/3}-0.86A^{-1/3})\,\mathrm{fm}$. The constant $n$ is fixed so that the distribution is normalized to 1. The other parameters in~(\ref{eq:ica}) take the same values as in~(\ref{eq:icp}). Therefore the standard nuclear transverse thickness function $T_A$ is the only new quantity introduced in the case of a nucleus target. To compute proton-nucleus cross sections we compute the yield at fixed impact parameters and integrate explicitly over $\bt$ in~(\ref{eq:defphi}). A problem with this approach is that at large impact parameters the saturation scale of the nucleus becomes too small for our formalism to be applicable. Therefore we impose $R_\text{pA}=1$ in the region where the saturation scale of the nucleus would be smaller than the one of the proton. This approach was first applied to single inclusive hadron production in Ref.~\cite{Lappi:2013zma} where it was shown to lead to values of the nuclear modification factor going to unity at large transverse momentum and compatible with LHC data.

\section{Results for $\Jpsi$ production}

To obtain the $J/\psi$ production cross section from the $c\bar{c}$ pair production cross section~(\ref{eq:dsigmaccbarcoll}), we use here the simple color evaporation model (CEM). A more rigorous way to treat hadronization could be to rely on non-relativistic QCD, as done in Ref.~\cite{Ma:2015sia}. In the CEM, we have
\begin{align} 
\frac{\ud\sigma_{\Jpsi}}{\ud^2\P_{\perp}\ud Y}
=
F_{\Jpsi} \; \int_{4m_c^2}^{4M_D^2} \ud M^2
\frac{\ud\sigma_{c\bar c}}
{\ud^2\P_{\perp} \ud Y \ud M^2}
\, ,
\label{eq:dsigmajpsi}
\end{align}
where $M$ is the invariant mass of the quark-antiquark pair and $m_c$ and $m_D$ are the charm quark and the $D$ meson mass, respectively. In the following we take $m_D=1.864$ GeV and we vary $m_c$ between 1.2 and 1.5 GeV. $F_{\Jpsi}$ is a non-perturbative constant related to the probability for a $c\bar{c}$ pair with an invariant mass lower than $2 m_D$ to hadronize into a $J/\psi$ meson and it cancels in the nuclear modification factor, therefore its exact value is not important here.

\begin{figure}[tb]
	\includegraphics[scale=1.15]{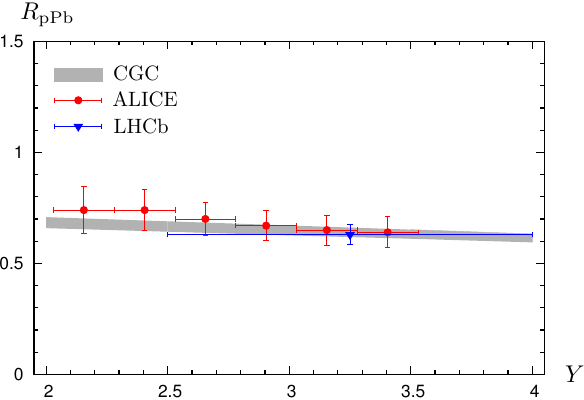}
	\caption{Nuclear modification factor for $\Jpsi$ production as a function of $Y$. Data from Refs.~\cite{Abelev:2013yxa,Aaij:2013zxa}.}
	\label{fig:RpA_Y_Jpsi}
\end{figure}

\begin{figure}[tb]
	\includegraphics[scale=1.15]{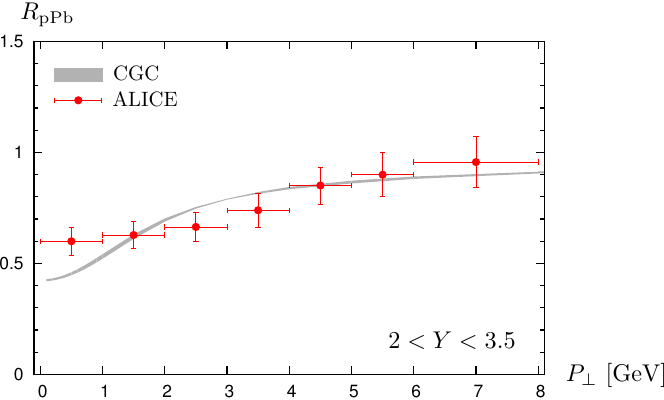}
	\caption{Nuclear modification factor for $\Jpsi$ production as a function of $P_\perp$ ($2<Y<3.5$). Data from Ref.~\cite{Abelev:2013yxa}.}
	\label{fig:RpA_pT_Jpsi}
\end{figure}

In Figs.~\ref{fig:RpA_Y_Jpsi} and~\ref{fig:RpA_pT_Jpsi} we show the nuclear modification factor for $\Jpsi$ production at $\sqrt{s_{NN}}=5$~TeV as a function of $Y$ and $P_{\perp}$ respectively. The uncertainty band contains the variation of $m_c$ between 1.2 and 1.5 GeV and of the factorization scale $Q$ between $M_\perp/2$ and $2M_\perp$ with $M_\perp=\sqrt{M^2+P_{\perp}^2}$. We see that our results are in much better agreement with ALICE and LHCb data than those shown in Ref.~\cite{Fujii:2013gxa}. This is mostly due to the fact that the initial saturation scale for the lead nucleus in our approach is significantly smaller than the simple scaling by $A^{1/3}$ compared to a proton. We note that the authors of Ref.~\cite{Fujii:2013gxa} have recently presented updated results~\cite{Fujii:2015lld} using a smaller initial saturation scale for the nucleus $Q_{s0,A}^2=\frac{1}{2} A^{1/3} Q_{s0,p}^2$ which also leads to a better agreement with data.

\section{Results for $D$ meson production}

From the $c\bar{c}$ pair production cross section~(\ref{eq:dsigmaccbarcoll}) we can also obtain the cross section for $D^0$ meson production as
\begin{align}
	\frac{\ud \sigma_{D^0}}{\ud^2 P_{\perp}\ud Y}=& Br(c \to D^0) \int \frac{\ud z}{z^2} D(z) \nonumber \\
	& \times \int \ud^2\qt \, \ud y_q \frac{\ud \sigma_{c\bar{c}}}{\ud^2\pt \ud^2\qt \ud y_p \ud y_q} ,
\end{align}
with $\pt=P_{\perp}/z$ and $y_p=Y$. We use the fragmentation function parametrization from~\cite{Kartvelishvili:1977pi}:
\begin{equation}
	D(z)=(\alpha+1)(\alpha+2)z^\alpha(1-z) ,
\end{equation}
with $\alpha=3.5$~\cite{Aaron:2008ac}.

\begin{figure}[tb]
	\includegraphics[scale=1.15]{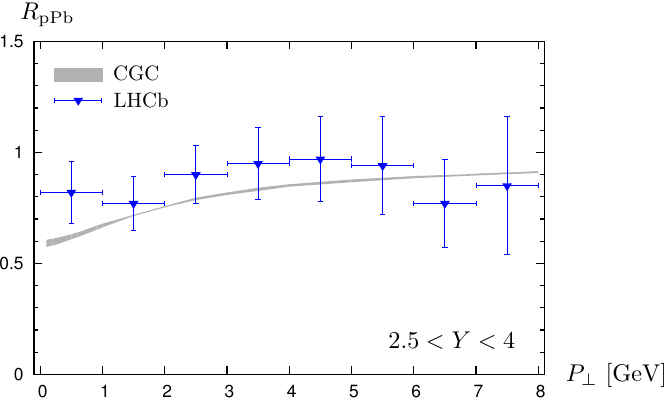}
	\caption{Nuclear modification factor for $D^0$ meson production as a function of $P_\perp$ ($2.5<Y<4$). Data from Ref.~\cite{LHCbConf}.}
	\label{fig:RpA_pT_D}
\end{figure}

In Fig.~\ref{fig:RpA_pT_D} we show the nuclear modification factor for $D^0$ meson production at $\sqrt{s_{NN}}=5$~TeV as a function of $P_{\perp}$. The uncertainty band is computed in the same way as in Figs.~\ref{fig:RpA_Y_Jpsi} and~\ref{fig:RpA_pT_Jpsi}. The conclusions are similar to the case of $\Jpsi$ suppression: the use of the optical Glauber model leads to less suppression and a better agreement with data than the first calculation of this process in the CGC formalism~\cite{Fujii:2013yja}.

\section{Conclusions}

In this work we studied the nuclear suppression of forward $\Jpsi$ and $D$ meson production in high energy proton-nucleus collisions in the color glass condensate formalism. To avoid introducing new parameters for the initial saturation scale of the nucleus, we used the optical Glauber model to relate the initial condition of a nucleus to the one of the proton, which is well constrained by HERA DIS data. This leads to less suppression and a better agreement with recent LHC data than the first study of these processes in this formalism.

\section*{Acknowledgements}
T.~L. and B.~D. are supported by  the Academy of Finland, projects 267321, 273464 and 303756 and by the European Research Council, grant ERC-2015-CoG-681707.
H.~M. is supported under DOE Contract No. DE-SC0012704.
This research used computing resources of 
CSC -- IT Center for Science in Espoo, Finland.



%% References

%\nocite{*}
%\bibliographystyle{elsarticle-num}
%\bibliography{biblio}






% \end{thebibliography}

\end{document}